# Hard-X-Ray Optics Development at Marshall Space Flight Center

Brian D. Ramsey and Martin C. Weisskopf

Space Sciences Department, NASA/Marshall Space Flight Center

The motivation for the development of hard-x-ray focusing optics, those operating at energies approaching 100 keV, is clear. Grazing incidence telescopes, such as that at the heart of the Chandra X-Ray Observatory[1] used for astronomical observations, have brought about spectacular advances at soft-x-ray wavelengths (< 10 keV), yet the hard-x-ray region, where such optics are yet to be routinely used, remains relatively unexplored at high sensitivity and fine angular resolution. The power of focusing, which concentrates source flux but not background, is such that even modest collecting areas can give a large increase in sensitivity over non-focusing devices.

The angle below which x-rays can be efficiently scattered from smooth surfaces, the critical angle, $\delta_c$, is quite small; hence the term grazing incidence. Away from absorption edges $\delta_c$ (deg) ~ (1/E [keV]) $\sqrt{\rho}$ (g/cm$^3$), where E is the x-ray energy in keV and $\rho$ is the density of the reflecting material (typically gold or iridium). For soft x-rays, around 1 keV, the critical angle is a few degrees, but it scales approximately inversely with energy and therein lies the challenge for hard-x-ray optics: as the graze angle decreases the projected area of a mirror shell becomes very small. To overcome this we must utilize a large number of mirror shells and as long a focal length as possible. For convenience, the shells can be nested in a mirror module, then multiple modules can be utilized each with its own focal plane x-ray detector. A typical mirror shell configuration first proposed by Wolter[2] is shown in Figure 1. Here two reflections are utilized, the first from a parabolic surface and the second from a hyperbolic surface, to give an image that is essentially coma free.

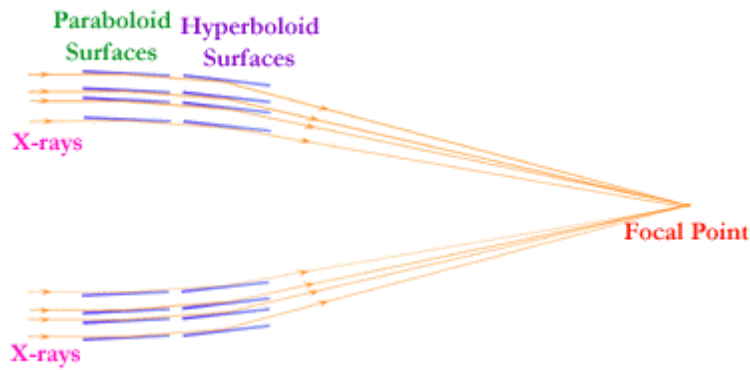

**Figure 1**. A Wolter-1 mirror configuration containing 4 nested shells

There has also been much discussion of coating the optics with graded multilayers, which, unlike conventional coatings, can exhibit moderate reflectivity at grazing angles several times the critical value. While much progress has been made with these coatings, they place stringent requirements on surface micro-roughness and require precise control of hundreds of very thin layers of material on the inside of each mirror shell, which make them unattractive, at least for the budget-limited balloon payload component of our research. Hard-x-ray telescopes with conventional coatings have certain advantages -- greater effective area per unit mass, less diffractive scattering by surface micro-roughness, less stringent manufacturing requirements -- over a multilayer-coated telescope[3].

The mirror fabrication process that we are developing is that of electroformed nickel replication (ENR). In this, nickel mirror shells are electroformed onto a figured and superpolished aluminum mandrel from which they are later released by differential thermal contraction. This process was pioneered in Italy for x-ray mirror fabrication and has been used for soft-X-ray astronomy in such missions as XMM-Newton[4], which featured 3 mirror modules each with 68 electroformed nickel shells.

A distinct advantage of the electroforming process is that the resulting mirror shells are full circles of revolution and thus are inherently very stable. This stability permits good figure accuracy, and hence very good angular resolution. A second advantage is that multiple identical copies can be made from a single mandrel and this permits the easy fabrication of multiple mirror modules. One drawback of the approach for future space applications, however, is the high density of nickel, which necessitates very thin shells for the lightweight optics necessary to keep costs reasonable. The shells, however, must be strong enough to withstand the stresses of fabrication and subsequent handling without being permanently deformed. They must also be electroformed in an ultra-low-stress environment to prevent stress-induced distortions once they are released. To ensure high-quality optics, we have therefore made developments in: 1) material strength; 2) adhesion and release, and; 3) plating bath stress control[5,6]

HERO, for High Energy Replicated Optics, is an evolutionary balloon program we are using as a test bed. It utilizes in-house-fabricated hard-x-ray mirrors plus detectors, gondola and pointing system. We utilize a large number of shallow-graze-angle, iridium-coated shells, moderately nested in multiple mirror modules. The use of a large shell length to diameter ratio reduces the number of mandrels. To keep costs appropriate to a balloon program, the mandrels are ground to sub-micron-accuracy figure and then polished, to 3-4 Å rms surface roughness, on simple in-house-designed machines. A multipart plater then permits several shells to be electroformed simultaneously. Figure 2 shows the electroforming tank (left), and a sample of the resulting mirror shells (right).

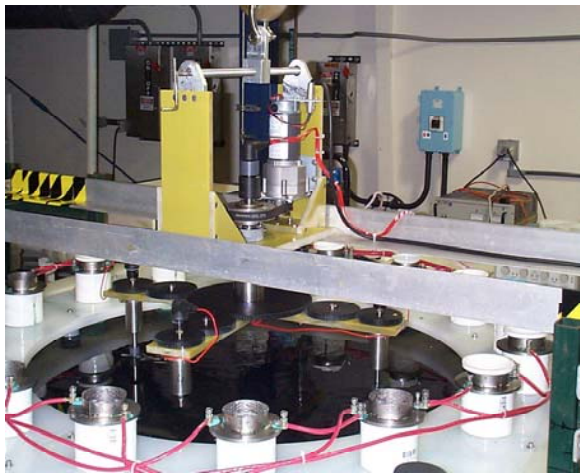 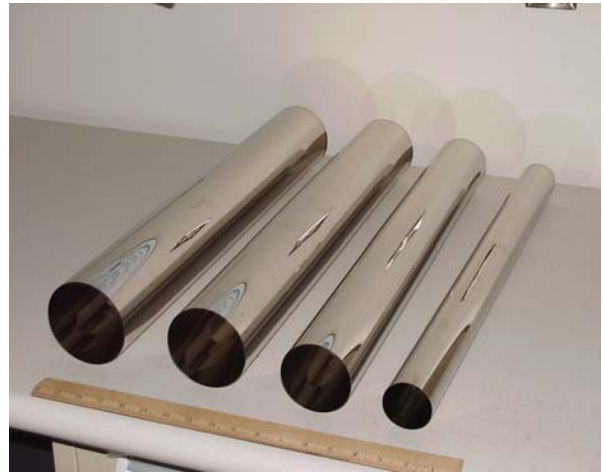

**Figure 2:** At left, the Ni/Co alloy electroforming bath with the multipart plater. At right, a selection of resulting mirror shells.

An early flight of a HERO prototype in May 2001 produced the first hard-x-ray images of astronomical objects and demonstrated sub-arc-minute angular resolution with six 3-m-focal-length mirror shells[7]. Since that time mirror fabrication techniques have been refined and a much larger payload, that will eventually feature 240 6-m-focal length mirrors, is under construction. These mirrors will be arranged in 16 housings each containing 15 nested shells ranging in diameter from 50 to 94 mm. A HERO partial payload flight is scheduled this spring from Fort Sumner, New Mexico. The payload consists of 8 mirror modules each with 8 mirror shells, giving a total effective area of 60 cm$^2$ at 40 keV. X-ray testing of these shells is being performed at MSFC, with the optics and test detector located in air at the end of a 100-m-long beam tube. Figure 3 shows a close up of a module with the 4 shells in Figure 2 mounted.

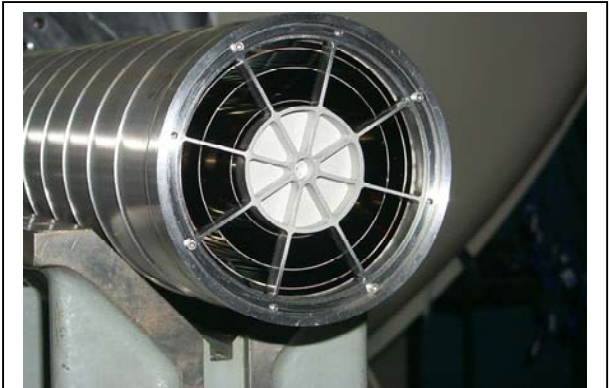

**Figure 3:** A test module with 4 shells mounted

One goal with the new balloon payload was to improve the mirror quality substantially over those originally flown in 2001. Typical metrology for current mandrels gives a performance prediction for the shells of around 8-10 arcsec half-power diameter. X-ray tests reveal shell performances in the 13-15 arcsec range with modules running around 17 arcsec. This module performance is about a factor of three better than our original units, but further improvements are possible. The mandrels themselves are conical approximation to a Wolter-1 geometry. This is done to simplify fabrication and thus reduce costs to those appropriate for a balloon program, but also limits the performance to a theoretical 6-10 arcsec HPD for the range of shell diameters. The difference between the mandrel predicted performance and the actual shell measurements are due to minute figure distortions in the shell fabrication process. We are investigating techniques for annealing out these effects. Finally, errors in the mounting of concentric shells can be overcome through an optical monitoring system that ensures that all shells are co-aligned and circular before bonding to the support spider takes place.  Thus we believe that even higher resolution optics are possible.

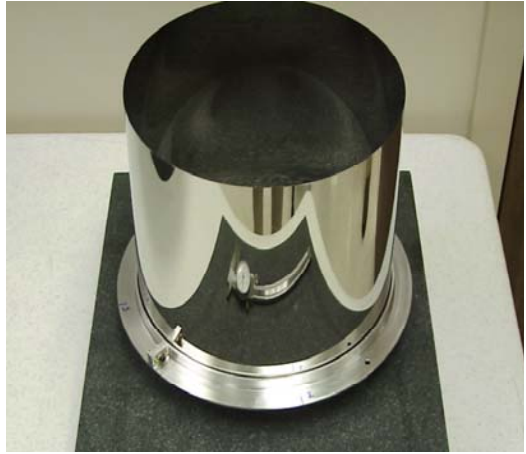

**Figure 4:** A 100-μm-thick test shell mounted in a support ring.

Another component of our ENR development work, in conjunction with the Smithsonian Astrophysical Observatory (SAO) and the Astronomical Observatory of Brera, Italy, concerns the building and testing of a prototype hard-x-ray telescope for possible use on the Constellation-X mission. Our role in this is the production of two demonstration shells, one 150 mm in diameter and coated with sputtered iridium, and the other 230 mm diameter to be coated with graded-multilayers. The challenging aspects of this work are the very thin mirror shell walls dictated by the tight Constellation-X weight budget, and the high-surface-quality finish on the interior of the shells designated for multilayer coating. The former necessitates production of shells of only 100 μm thickness, and the latter dictates a surface finish of better than 3 Å rms.

Test shells as shown in Figure 4 have been fabricated to demonstrate the viability of 100-micron-thick shell production and to evaluate the effects of coating stresses on the shell figure. Measurements on these shells indicates that multilayer stresses do not produce measurable figure distortions and that simple support rings can permit handling and mounting of ultra-thin mirror shells while maintaining their shape.

Both Constellation-X mandrels have been completed and are ready for electroforming. Wyko data indicate surface roughnesses of 2.7 Å for the larger mandrel and a figure error of less than 0.1 micron. The performance prediction for shells from this mandrel, a conical approximation, is 10 arcsec HPD. Test shells from these mandrels will be fabricated later this year.

In conclusion we believe that electroformed nickel replication offers an extremely attractive and inexpensive solution to the problem of hard-x-ray optics production. The process lends itself readily to the multiple mirror module approach that small graze angles necessitate and the resulting shells provide excellent angular resolution that results in observations at high sensitivity. Finally, with the use of high-strength alloys one can achieve the stringent weight budgets that future missions require.